\documentclass[preprint,showpacs,showkeys,amsmath,amssymb,superscriptaddress,nofootinbib]{revtex4}

\usepackage{graphicx}
\usepackage{epsfig}
\usepackage{dcolumn}
\usepackage{bm}

\begin{document}

\title{Characteristic feature of self-consistent mean-field in level crossing region}

\author{Lu Guo}
\email{guolu@mcs.ibaraki.ac.jp}
\affiliation {Department of Mathematical Science, Ibaraki University, Mito, Ibaraki 310-8512, Japan}
\author{Fumihiko Sakata}
\affiliation {Department of Mathematical Science, Ibaraki University, Mito, Ibaraki 310-8512, Japan}
\author{En-guang Zhao}
\affiliation {Institute of Theoretical Physics, Academia Sinica, P.O. Box 2735, Beijing 100080, China}
\date{\today}

\begin{abstract}
A shape change of the self-consistent mean-field induced by a configuration change is discussed within the conventional constrained Hartree-Fock (CHF) theory. 
It is stressed that a single-particle level crossing dynamics should be treated carefully,  because the shape of the mean-field in such a finite many-body system as the nucleus strongly changes depending on its configuration. 
This situation is clearly shown by applying an adiabatic assumption, where the most energetically favorable single-particle states are assumed to be occupied.
The excited HF states and the continuously-connected potential energy curves are given by applying the configuration dictated CHF method. The effect of pairing correlation is discussed in the level crossing region.
Triaxial deformed results in our Hartree-Fock-Bogoliubov (HFB) calculation with Gogny force nicely reproduce the available experimental data of Ge isotopes.
From our numerical calculation, it is concluded that the CHFB state is more fragile than the CHF state in the level crossing region.
\end{abstract}

\maketitle

\section{\label{level1} Introduction}
The HFB theory has been regarded to give a good starting point for a quantum description of an isolated many-fermion system like a nucleus.
It defines an approximate ground state as well as a corresponding set of single-particle states, which have been used for the shell model calculations or the RPA treatments.
Paying a price for breaking various conservation laws satisfied by the Hamiltonian, one obtains such an appropriate mean-field that might incorporate various correlations of the system as much as possible. 
Its many important theoretical concepts like a stability of the mean-field, a restoration dynamics of the broken symmetries, etc constitute an outstanding virtue of the self-consistent mean-field.

Owing to intensive studies on the high-spin physics such as the band termination, band crossing, shape coexistence, identical bands and superdeformed bands, it has been shown that an introduction of many mean-fields to one nucleus successfully explains various excited states appearing near the yrast region. 
Namely, many mean-fields specified by an approximate angular momentum, aligned angular momentum, single-particle configuration and other approximate quantum numbers, provide us with physical, simple and satisfactory understanding for a number of excited states better than a full quantum mechanical treatments.
A rapid expansion of applications of the mean-field methods both with \cite{ref1,ref2,ref3,ref4,hf1,hf2,RWL02} and without \cite{JL92, RB89,TB,RB} self-consistency gives rise to various important theoretical issues in the microscopic theory of many-body system.
Among others, a property change in the single-particle states and its dynamical relation to the mean-field, a stability of the excited mean-field, a restoration dynamics of broken symmetries for the excited mean-fields and a dynamics of the single-particle level crossing are interesting subjects to be explored.

In the level crossing region, there have been made many discussions on an applicability of the cranked mean-field theory \cite{Hama76}, because there appears not only a large angular momentum fluctuation, but also a spurious interaction between two crossing orbits not at a given angular momentum but at a given angular frequency.
In order to approximately restore the angular momentum conservation law by hand near the level crossing region, at an expense of the virtue of the mean-field theory stated above, it might be meaningful to eliminate the spurious interaction and to introduce a diabatic orbit \cite{TB89}. 
With regards to the non-adiabatic (diabatic) effects due to the fast nuclear rotation,  rotation-induced time-odd components of one-body density have been introduced and analysed in Refs. \cite{odd1, ref1}, and detailed analysis \cite{ref1,ref2,ref3,ref4} shows their importance in the identical and superdeformed bands.
A dynamical and quantum origin of these diabatic or non-adiabatic effects appeared in the mean-field theory also exhibits one of the most interesting problems in the theory of nuclear collective motion.

Since these interesting problems have been mainly discussed qualitatively, it is necessary to study more deeply what actually happens in the self-consistent mean-field at the level crossing point, and how strongly it undergoes a change when it acquires additional deformation, additional angular momentum and when its single-particle configuration changes. 
Aiming to understand microscopically how the mean-field changes by itself as a smooth function of global variables, a new numerical method called the configuration dictated CHF method has been proposed in Ref. \cite{CD1} and recently applied for the rotational bands \cite{CD2,CD4}. 
In contrast with the conventional numerical method to solve the CHF equation, which are mainly applied on a selected set of mesh points on the global variables, our method intends to keep an identity of the CHF states and the calculation is made in a small step so that the resultant solutions are regarded as a continuous function of the global quantities like quadrupole deformation, angular frequency, etc.  
Applying this method,  one may construct various configuration-dependent mean-fields as a function of rotational frequency or nuclear deformation, and explore what happens at the level crossing point. 

In this paper, we will pay special attention how strongly the self-consistent mean-field is affected by the configuration change, which is expected to occur at the level crossing region.
This situation is well studied by comparing two different numerical calculations: one is the adiabatic mean-field theory and the other is the configuration dictated approach. By using the CHF and CHFB calculations the low-lying properties of Ge isotopes will be discussed in terms of the relation between high-$j$ intruder occupation and deformation, which has been studied in the deformed mean-field theory \cite{JL92,WN85,JD88} and shell model calculations \cite{KL03NPA,AP02} in the A$\sim$80 mass region.

This paper is organised as follows. 
Section \ref{level2} gives a brief outline of the theoretical method. 
In Sec.~\ref{level3}, a specific feature of the adiabatic mean-field theory is discussed and
low-lying excited states and continuously-connected potential energy curves are given by the configuration dictated method for Ge isotopes. The effect of pairing correlation near the level crossing region and a fragility of the CHF and CHFB mean-fields are discussed based on the numerical calculation.
Triaxial deformed results in our HFB calculation with Gogny force are compared with the available experimental data.
Section \ref{level4} is devoted to discussion and summary.  

\section{\label{level2} Theoretical method}
In this section, we discuss the necessary formulae and methods for analysing structure changes near level-crossing region.
For the sake of simplicity, we will discuss them within the constrained HF rather than the constrained HFB formalism.
Inclusion of the pairing correlation is straightforward.

In the present work, the self-consistent CHF(B) equation is solved using three dimensional harmonic oscillator basis with the Gogny D1S interaction \cite{Gog1,Gog2,Gog3,Gog4,Gog5,Gog6}. 
The Coulomb interaction and the center of mass correction up to the exchange terms are also taken into account. 
To save the CPU time in numerical calculation, we impose the $\hat{P}e^{-i\pi\hat{J_z}}$ (z-simplex) and
$\hat{P}e^{-i\pi\hat{J_y}}\hat{\tau}$ ($\hat{S_y ^T}$) symmetries \cite{JD1,JD2}, where $\hat{P}$ is the parity operator, $e^{-i\pi\hat{J_i}}$ the rotation operator around {\it i} axis by angle $\pi$, and $\hat\tau$ the time reversal operator. 
Due to the z-simplex and $\hat{S_y ^T}$ symmetries, a mass asymmetry of the nucleus is allowed only along the x axis. 
To keep the center of mass motion fixed, we impose a quadratic constraint in the x-axis direction. 
The constrained HF equation to be solved is thus given as
\begin{equation}
\delta\biggl(\langle \hat{H} \rangle+\frac{1}{2}C\bigl(\langle\hat
Q_{20}\rangle-\mu\bigr)^2+\frac{1}{2}\alpha_x \langle\hat x \rangle ^2\biggr) =0 ,
\label{eq:chf1}
\end{equation}
with constraint
\begin{equation}
\langle\hat Q_{20}\rangle = q , \quad
\langle\hat x\rangle = 0 .
\end{equation}
A quadratic constraint of quadrupole moment is used to treat concave areas of the energy surface. $\mu$ is an input parameter which allows us to vary the expectation value $\langle\hat Q_{20}\rangle$. 
Meaning of $\it{C}$ was discussed in Ref. \cite{HFL}, and is chosen to be $1.0*10^{-3}[{\textrm {MeV}}/{\textrm {fm}}^4]$ in our calculation. 
The Lagrange multiplier $\lambda$ is given by
\begin{equation}
\lambda = C\bigl(\mu-\langle\hat Q_{20}\rangle\bigr) = dE/dq,  
\quad
E\equiv \langle\hat H \rangle,
\end{equation}
and an effective value of $\lambda$ is allowed to change during the iteration.
In our numerical calculation, we take $\alpha_x=1.0*10^{-4}[{\textrm {MeV}}/{\textrm {fm}}^2]$.

For convenience, the constrained Hamiltonian is denoted by $\hat C\equiv \hat H-\lambda \hat Q_{20}$, hereafter.
Equation (\ref{eq:chf1}) is then expressed as (for simplicity, the center of mass correction constraint will be 
omitted in the following formula)
\begin{equation}
\delta\langle\psi(q)|\hat C |\psi(q)\rangle=0.
\end{equation}
 Hereafter an explicit deformation dependence of the CHF state $|\psi(q)\rangle$ is used instead of $|\rangle$. 
For $|\psi(q)\rangle$, the constrained Hamiltonian is expressed as
\begin{eqnarray}
\hat C(q) & \equiv & \hat H-\lambda(q)\hat Q_{20} \nonumber \\
& =& \langle\psi(q)|\hat C(q)|\psi(q)\rangle +\sum_{\mu}\epsilon_{\mu}(q)\hat c_\mu ^\dag(q)\hat c_\mu(q)
-\sum_{i}\epsilon_{i}(q)\hat c_i (q) \hat c_i ^\dag(q) + :\hat C(q): ,
\end{eqnarray} 
where ${\hat c_{\mu}^\dag(q)}$ and ${\hat {c_i}^\dag(q)}$ are particle- 
and hole-creation operators satisfying
\begin{equation}
\hat c_{\mu}(q) |\psi(q)\rangle = \hat c_i^\dag (q)|\psi(q)\rangle =0.
\end{equation}
Here and hereafter, the particle states are denoted by $\mu$, $\nu$ and hole states by $i$, $j$. If we do not distinguish, we use the letters $k$, $l$. 
The operator $:\hat C(q):$ denotes the two-body residual interaction consisting of the normal-ordered product of four fermions in the two-body interaction with respect to $|\psi(q)\rangle$. 

To understand the microscopic dynamics responsible for the origin of the quadrupole deformation, which will be discussed in the next section, the quadrupole operator is expressed as
\begin{equation}
\hat Q_{20}=q +\sum_{\mu i}\{Q_{\mu i}(q)\hat c_{\mu}^\dag\hat c_i+{\textrm {h.c.}}\} 
+\sum_{\mu \nu}Q_{\mu \nu}(q)\hat c_{\mu}^\dag\hat c_\nu
-\sum_{ij}Q_{ij}(q)\hat{c_j} \hat c_i ^\dag .
\label{eq: Q20}
\end{equation}
Here h.c. denotes the Hermitian-conjugation of the former term.
The deformation-dependent particle-hole, hole-hole and particle-particle components  $Q_{\mu i}(q)$,  $Q_{i i}(q)$ and  $Q_{\mu \mu}(q)$ indicate how a microscopic structure of the self-consistent mean-field $|\psi(q)\rangle$ changes, and what kind of single-particle configuration is favored as a function of nuclear deformation. 
The quadrupole deformation of the system is given by
\begin{equation}
q=\sum_{i=1}^{{\textrm {N}}} Q_{ii}(q),
\end{equation}
$N$ being the number of nucleons.

In the present work, we solve CHF(B) equation by two ways. One is the conventional adiabatic method as will be discussed in Sec.~\ref{level31}. 
In this case, each iterative process is achieved by introducing a new density matrix constructed by the first N-lowest eigenvectors (adiabatic assumption).
A convergence is considered to be completed when the resultant density matrix is equivalent to the preceding one within a given accuracy.
In this way, the most energetically favorable CHF(B) state satisfying a given constrained condition is obtained. 
In this method, one does not pay any attention to mutual relations between two neighbouring Slater-determinants with slightly different quadrupole deformation $|\psi(q)\rangle$ and $|\psi(q+\Delta q)\rangle$. 

To understand how the CHF(B) state undergoes a structure change depending on the quadrupole deformation, it is desirable to obtain $|\psi(q+\Delta q)\rangle$ in such a way that it can be regarded as a smooth function of $\Delta q$.  
For this aim, we apply the second method called the configuration dictated method, which is briefly recapitulated below.
Let $|\psi(q)\rangle$ be a known CHF state satisfying a condition $\langle \psi(q)| \hat Q_{20} |\psi(q)\rangle =q$. 
To find a new CHF state $|\psi(q+\Delta q)\rangle$ which is supposed to be continuously connected with $|\psi(q)\rangle$, we exploit a condition
\begin{equation}
\lim_{\Delta q \rightarrow 0}\langle\varphi_i(q)|\varphi_j(q+\Delta q)\rangle=\delta_{i,j},\quad i, j=1,\cdots,N,
\label{cdm}
\end{equation}
where $\{|\varphi_i(q)\rangle, i=1,\cdots,N\}$ denotes a set of occupied wave functions constructing the single Slater determinant $|\psi(q) \rangle$.
That is to say, each small increment $\Delta q$ is numerically adjusted by the maximum overlap criterion (\ref{cdm}) under a given accuracy, so as to maintain the identity of CHF states.
In our calculations, $\Delta q$ is so determined as to fulfil the condition 
\begin{equation}
|\langle\varphi_i(q)|\varphi_i(q+\Delta q)\rangle|^2 > 0.9,\quad i=1,\cdots,N. 
\label{cdm9}
\end{equation}
In this way, the configuration specifying $|\psi(q)\rangle$ is considered to be kept continuously as a function of $q$.
Since the CHF state at $q+\Delta q$ is dictated by the configuration of the preceding CHF state at $q$ rather than the adiabatic assumption, this method is called the configuration dictated CHF method. 

This method is also generalised to get an excited HF state.
Suppose there exists a HF state denoted by $|\phi_0\rangle$.
One may then introduce a certain single Slater determinant $|\phi^{RS}\rangle$, which is a simple $np-nh$ state with respect to $|\phi_0\rangle$ and is called a reference state (RS).
Here, it should be noted that the RS does not necessarily satisfy the HF condition.
To get the HF solution $|\phi_{\mbox{np-nh}}\rangle$ locating at the nearest distance from $|\phi^{RS}\rangle$, one may apply the following procedure in each iteration.

\begin{enumerate}
\item Start the HF iterative procedure with the RS as an initial single Slater determinant $|\phi_{\mbox{np-nh}}^{n=0}\rangle=|\phi^{RS}\rangle$.
\item Suppose the $n$th single Slater determinant $|\phi_{\mbox{np-nh}}^{n}\rangle$ is composed of a set of single-hole states $\{ |\varphi_i^{n}\rangle,i=1,\dots,N\}$, where $n$ denotes the number of iteration.
Here, it should be noticed that $\{ |\varphi_i^{n}\rangle,i=1,\dots,N\}$ does not necessarily consist of the first $N$ lowest states, but the occupied orbits. 
Using $|\phi_{\mbox{np-nh}}^{n}\rangle$, one may define the $(n+1)$th one body density $\rho_{\mbox{np-nh}}^{n+1}$, and the $(n+1)$th single-particle Hamiltonian $\hat h(\rho_{\mbox{np-nh}}^{n+1})$.
\item With the aid of $\hat h(\rho_{\mbox{np-nh}}^{n+1})$, one may get a set of single particle states $\{|\varphi_\alpha^{n+1}\rangle, \alpha=1,\dots,N,\dots \}$. 
In constructing the $(n+1)$th single Slater determinant $|\phi_{\mbox{np-nh}}^{n+1}\rangle$, one has to select $\{
|\varphi_i^{n+1}\rangle, i=1,\dots,N\}$ out of $\{|\varphi_\alpha^{n+1}\rangle,\alpha=1,\dots,N,\dots\}$.   
To maintain a property of $|\phi^{RS}\rangle$ at each iteration, one has to find a new set of occupied orbits $|\varphi_j^{n+1}\rangle$ in such a way that there holds the following conditions
\begin{equation}
|\langle\varphi_j^{RS}| \varphi_j^{n+1}\rangle|^2 >
|\langle\varphi_j^{RS}| \varphi_k^{n+1}\rangle|^2  , \quad
 \mbox{for all   } k > j \quad \mbox{with  } j=1,2,\cdots,N.  \label{cond}
\end{equation}
\item Iteration of the second and third steps should be repeated till $\{|\varphi_i^{n}\rangle,i=1,\dots,N,\dots\}$ and $\{|\varphi_i^{n+1}\rangle,i=1,\dots,N,\dots\}$ become equivalent within the required accuracy.
 When the iteration converges, one gets the excited HF state $|\phi_{\mbox{np-nh}}\rangle$ whose
configuration is dictated by that of $|\phi^{RS}\rangle$.
\end{enumerate}

The above-stated method is called a reference state (RS) method, and enables us to obtain many excited HF states.
Applying the configuration dictated method by starting with various excited HF states, one may get many CHF lines (i.e. many potential energy surfaces as a smooth function of the deformation $q$) formed by continuously-connected CHF solutions, which will be discussed in Sec.~\ref{level32}.

In the following calculation, both the adiabatic and configuration dictated CHF(B) are done in such a small step as the solution is considered to be a continuous function of the quadrupole deformation. 
Such a point-by-point heavy calculation is needed for discussing the dynamical change of nuclear system and the structure change of configuration-dependent mean-field as a continuous function of deformation.

The single-particle wave functions are expanded in a three-dimensional harmonic oscillator basis up to the principal quantum number $N_0=8$. 
The range parameters of Hermite polynomials have been optimised for each nucleus to reproduce the largest ground state binding energy. 
The optimised range parameters thus obtained include some effects of higher major shells. Note that, no optimisation for each nucleus has been done in the most HF and HFB calculations \cite{ref1,ref2,ref3,ref4}, although a larger configuration space is adopted.
It is well known that the optimal parameters change depending on $N_0$, and then become rather stable as the shell number $N_0$ increases \cite{HFL}.
In order to examine a size effect of the configuration space, low-lying spectra obtained by the RS method with $N_0=12$ are compared to those with $N_0=8$ for Ge isotopes, where the intruder orbit $g_{9/2}$ plays an important role and  provides us with an excellent opportunity to study the shell structure and shape coexistence. The good agreement for excitation energies and deformations between $N_0=8$ and 12 cases (see below) indicates a reliability of our discussion with $N_0=8$.

In our calculation, one set of range parameters has been used for each nucleus in numerically obtaining the CHF(B)-lines and excited states. 
A freedom of variation in the range parameters may lower the binding energy of CHF(B) states, which could lead to the effective mixing of different configurations. However, the use of fixed range parameters allows one to trace an evolution of the ground state configuration as a function of deformation, which makes the single-particle level crossing dynamics transparent.

In our numerical calculation, a convergence condition is given as
\begin{equation}
\sum_{k}\left|\epsilon^{(n)}_k(q)-\epsilon^{(n-1)}_k(q)\right|\leq 0.1 {\textrm [KeV]},
\label{eq12}
\end{equation}
where $\epsilon_k^{(n)}$ is s.p. energy in the $n$-th iteration.

\section{\label{level3} Numerical calculations}
\subsection{\label{level31} Adiabatic constrained HF}

\begin{figure}
\epsfxsize=8.0cm
\centerline{\epsffile{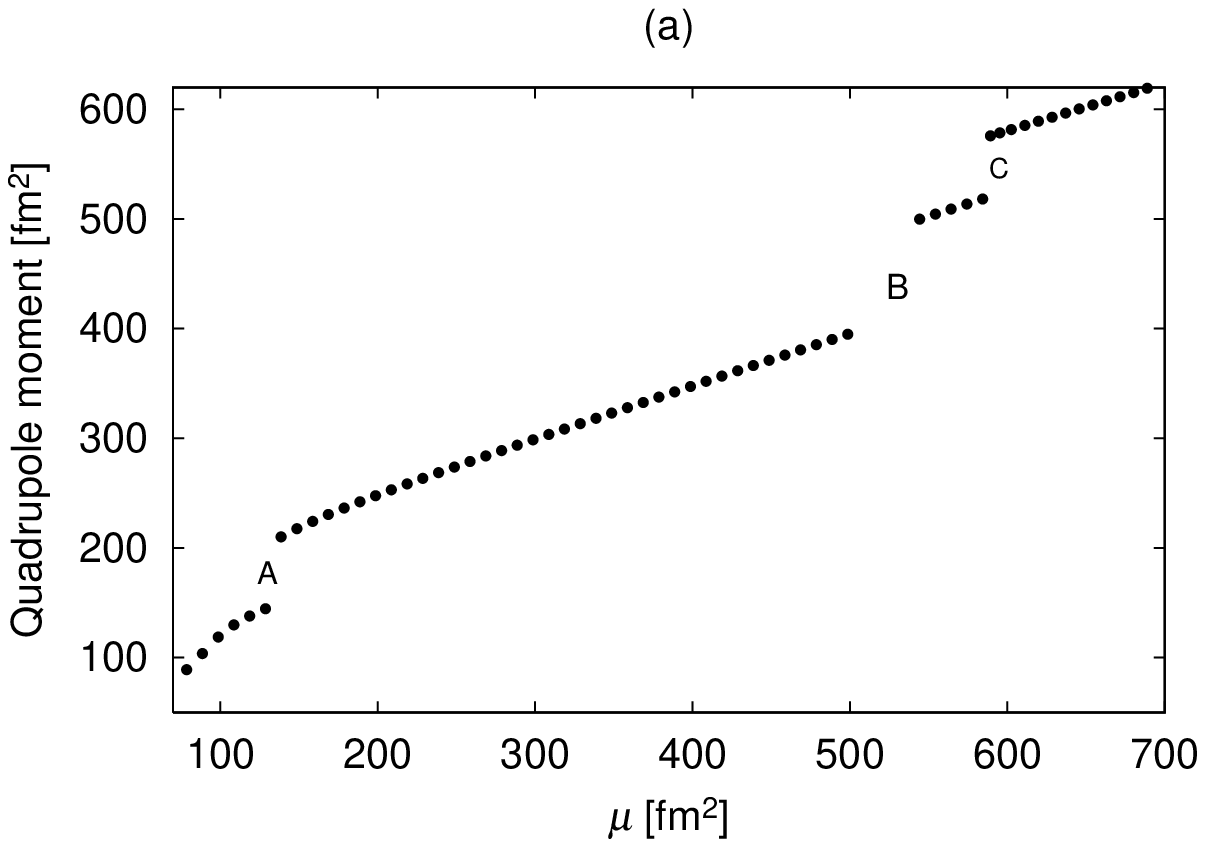}}
\centerline{\epsffile{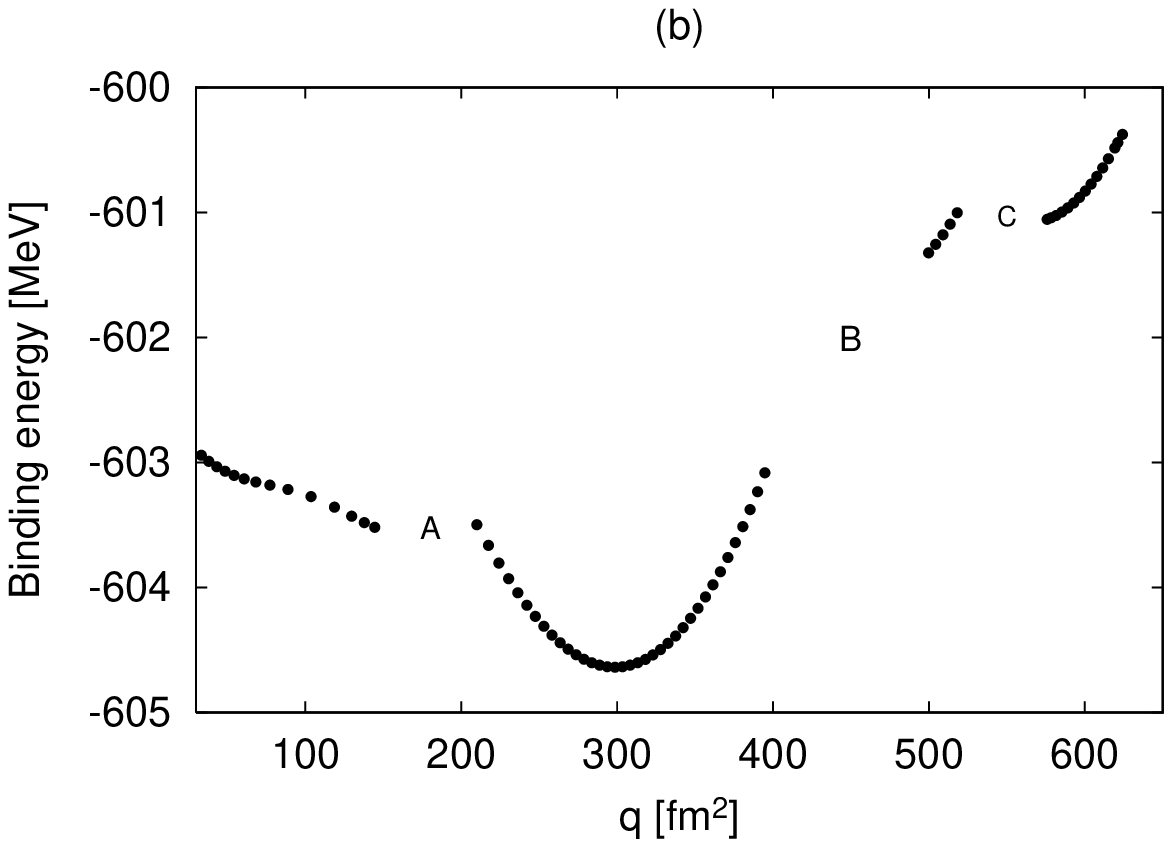}}
\caption{\label{fig:adi1} Adiabatic CHF for
$^{70}{\textrm {Ge}}$: (a) the calculated quadrupole moment as a
function of input quadrupole moment parameter $\mu$; (b) binding
energy as a function of the calculated quadrupole moment.}  
\end{figure}

The ground state for nucleus $^{70}{\textrm {Ge}}$ is obtained after an optimisation on triaxial deformation parameters of the Hermite polynomials. 
Starting from the ground state, the adiabatic CHF calculations are carried out for nucleus $^{70}{\textrm {Ge}}$ shown in Fig.~\ref{fig:adi1}.
As seen from Fig.~\ref{fig:adi1}(a), the resultant quadrupole moment changes smoothly as a function of input quadrupole moment parameter $\mu$, except for three points denoted by A, B and C. 
The binding energy as a function of the quadrupole moment also exhibits drastic change at these three points, which is observed in Fig.~\ref{fig:adi1}(b). 

\begin{figure}
\epsfxsize=8.0cm
\centerline{\epsffile{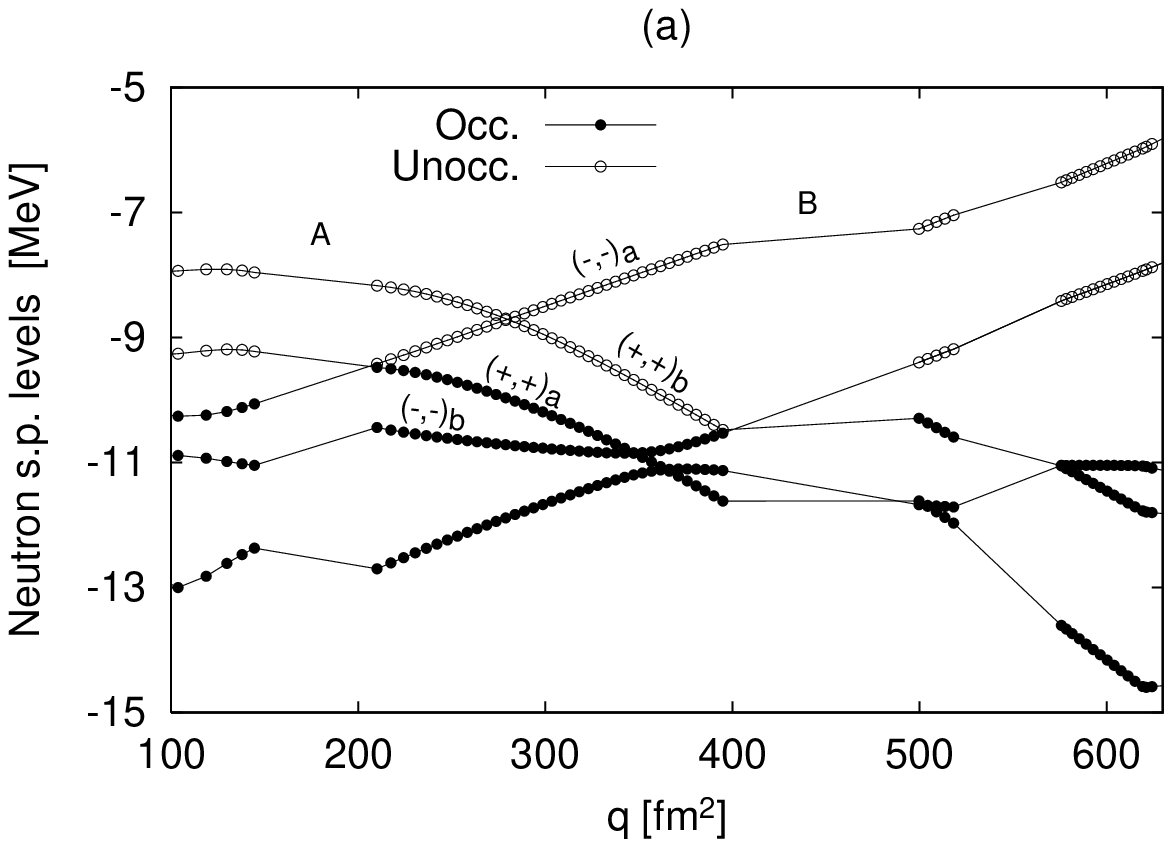}}
\centerline{\epsffile{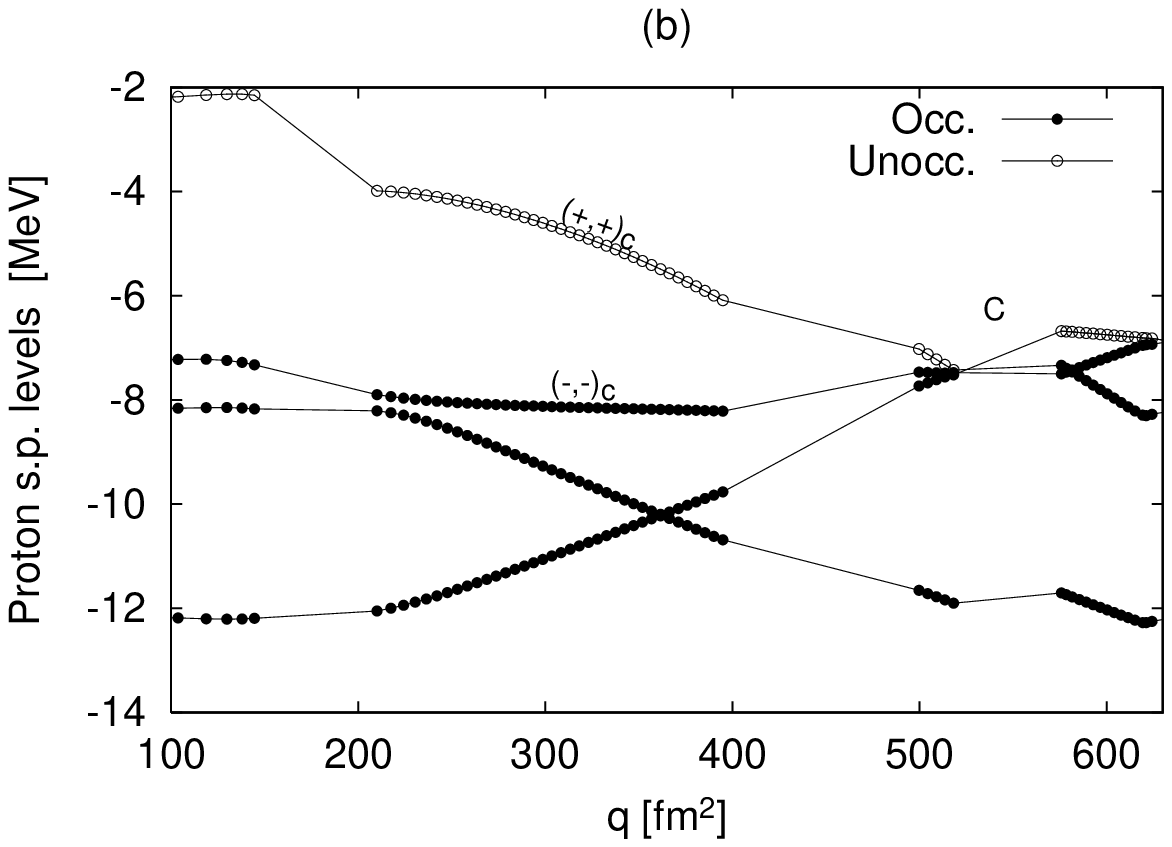}}
\caption{\label{fig:adi2} Neutron (a) and proton (b) single-particle energies near the Fermi surface for $^{70}{\textrm {Ge}}$. Occ. and Unocc. stand for occupied and unoccupied orbits in adiabatic CHF calculation, respectively. ($\pi$,$\alpha$) denotes the parity and signature, and its subscripts a, b and c represent the orbits responsible for the gaps A, B and C.} 
\end{figure}

To understand what happens in these three points from the microscopic point of view, the neutron and proton single particle energy levels are depicted in Fig.~\ref{fig:adi2} as a function of quadrupole moment $q$. 
Here Occ. and Unocc. are used to denote the occupied and unoccupied single particle orbits in adiabatic calculation.
They are specified by the parity and signature quantum numbers $(\pi,\alpha)$, because the asymptotic Nilsson quantum numbers are not good quantum numbers when the reflection symmetry is absent.
For convenience, a subscript a, b and c for $(\pi,\alpha)$ is used to identify the orbits responsible for the gaps A, B and C. 
In this mass region, $g_{9/2}$ orbit plays an important role for the shell structure and nuclear deformation. 
In the following discussion, the configuration characterising the CHF state is labelled as $[p_1 p_2, n_1 n_2]$, where $p_1$ ($n_1$) denotes the number of protons (neutrons) in the oscillator shell $N_0=3$, and $p_2$ ($n_2$) the number of protons (neutrons) in $g_{9/2}$ shell. 

The ground state configuration of $^{70}{\textrm {Ge}}$ is characterised by [12 0, 16 2], which is observed from Fig.~\ref{fig:adi2}. 
In Fig.~\ref{fig:adi2}(a), it is shown that the neutron unoccupied orbit $(-,-)_a$ belonging to $f_{5/2}$ or $p_{1/2}$, and the occupied orbit $(+,+)_a$ belonging to $g_{9/2}$ come close, as the quadrupole moment decreases from the ground state, and eventually cross with each other.
Although $\mu$ is equally discretized just before and after the gap A, the calculated quadrupole moment decreases dramatically.
After the gap A, the orbit $(-,-)_a$ becomes occupied and the orbit $(+,+)_a$ unoccupied in accordance with the adiabatic
assumption. 
Namely,  two neutrons are excited from $g_{9/2}$ to $f_{5/2}$ or $p_{1/2}$ relative to the ground state, forming a new excited configuration [12 0, 18 0].
 
In other two gap regions, the same microscopic structure change takes place. 
At the point B, the neutron orbit $(+,+)_b$ belonging to $g_{9/2}$ crosses with the orbit $(-,-)_b$ belonging to $f_{5/2}$. 
After the crossing B, two neutrons occupying $f_{5/2}$ move to $g_{9/2}$, forming an energetically favorable configuration [12 0, 14 4]. 
Between B and C, the adiabatic CHF solution keeps its identity until the proton orbit $(+,+)_c$ belonging to $g_{9/2}$, and $(-,-)_c$ belonging to $f_{7/2}$ or $p_{3/2}$ cross with each other. 
After the crossing, another configuration [10 2, 14 4] is realized and the third gap C appears.

A drastic change of the quadrupole moment just before and after the level crossing comes from three aspects. 
First is a smooth contribution on the quadrupole moment $q$ due to a change of the parameter $\mu$, which is expressed by a straight line segment between A and B in Fig.~\ref{fig:adi1}(a). 
Second is coming from a change of the configuration.
Since the quadrupole moment of the system is produced by a sum of the quadrupole moment of the occupied orbits, and since the configuration change takes place between two orbits with different deformation character, i.e., one is usually of deformation driving and the other is of deformation anti-driving, the configuration change between these orbits induces a substantial deformation change in such a finite many-body system as nucleus.
Third is the contribution from the rearrangement of mean-field due to the configuration change. 

To demonstrate to what extent each contribution devotes to the gap,
in the following, the above-stated three contributions are analysed quantitatively for the gap A. 
In this case, the change of quadrupole moment within an interval $\Delta\mu$=10.0 ${\textrm {fm}}^2$ is $\Delta q$=65.50 ${\textrm {fm}}^2$, whose proton and neutron contributions are $\Delta q_p$=20.08 ${\textrm {fm}}^2$ and $\Delta q_n$=45.42 ${\textrm {fm}}^2$, respectively. 
The gradient of the line segment between A and B shown in Fig.~\ref{fig:adi1}(a) is estimated as follows; $\mu_B-\mu_A$=360.0 ${\textrm {fm}}^2$ and $q_B-q_A$=184.7700 ${\textrm {fm}}^2$.
The proton contribution to the latter is $(q_B-q_A)_p$=81.3098 ${\textrm {fm}}^2$ and the neutron contribution is $(q_B-q_A)_n$=103.4602 ${\textrm {fm}}^2$. 
The gradient of the line segment between A and B is then given as: $(q_B-q_A)/(\mu_B-\mu_A)$=0.51325, whose proton and neutron contributions are  $(q_B-q_A)_p/(\mu_B-\mu_A)$=0.22586 and $(q_B-q_A)_n/(\mu_B-\mu_A)$=0.28739, respectively. 
With the aid of the gradient between A and B, one may estimate the first contribution as $\Delta q_p^{(1)}$=2.2586 ${\textrm {fm}}^2$ and $\Delta q_n^{(1)}$=2.8739 ${\textrm {fm}}^2$ within the interval $\Delta \mu$ in gap A. 
In other words, the total quadrupole moment only decreases $\Delta q^{(1)}$=$\Delta q_p^{(1)}+\Delta q_n^{(1)}$=5.1325 ${\textrm {fm}}^2$ when there happens no level crossing between occupied and unoccupied orbits.
This smooth effect corresponds to 11.25\% ($\sim$2.2586 ${\textrm {fm}}^2$) and 6.33\% ($\sim$2.8739 ${\textrm {fm}}^2$) of $\Delta q_p$ and $\Delta q_n$, respectively. 

\begin{figure}
\epsfxsize=8.0cm
\centerline{\epsffile{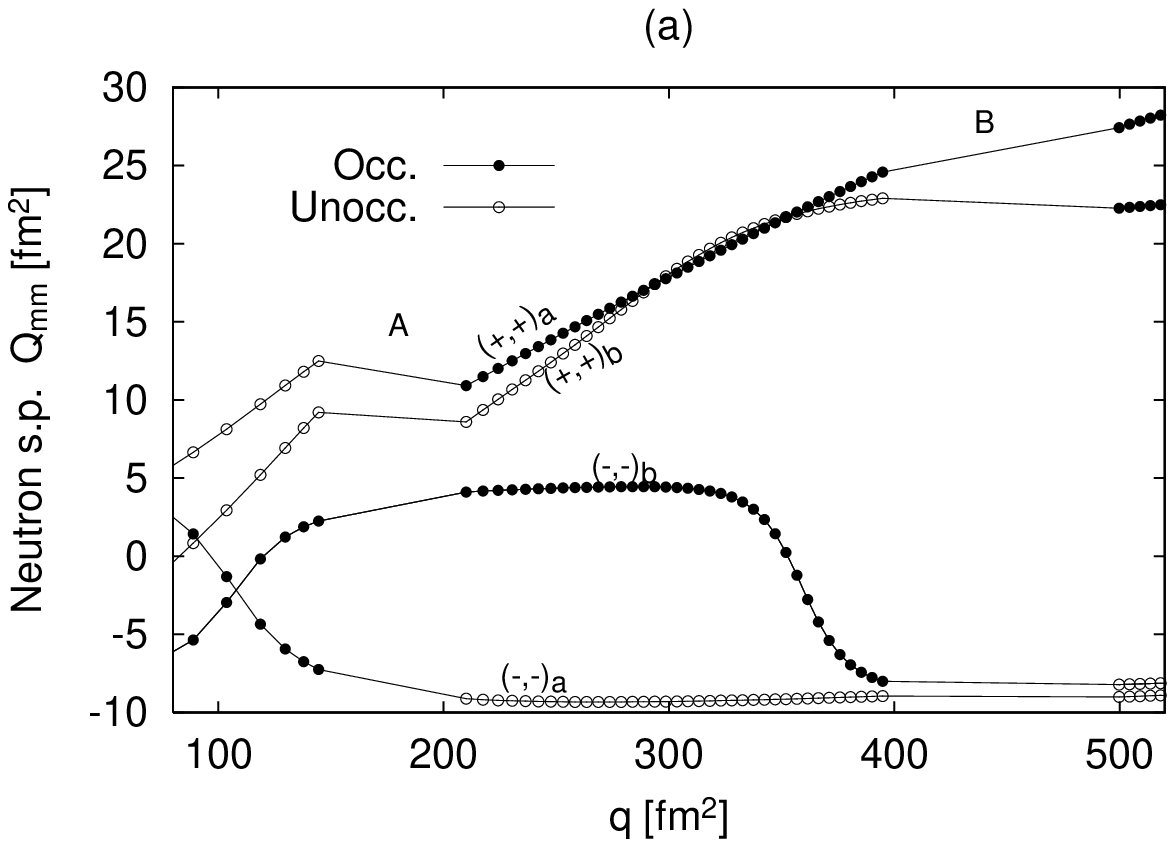}}
\centerline{\epsffile{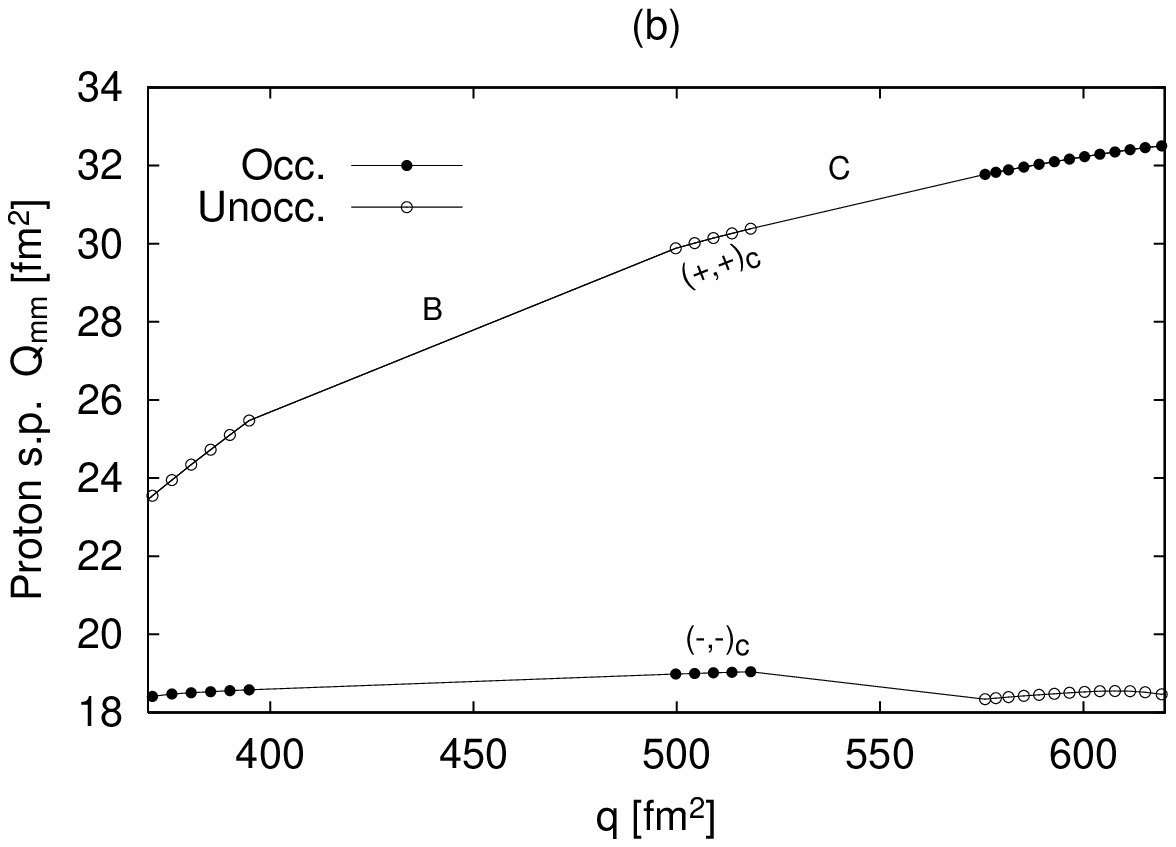}}
\caption{\label{fig:adi3} Neutron (a) and proton (b) diagonal matrix elements of single particle quadrupole moment for the specific crossing orbits responsible for the gaps A, B and C. The symbol is same as that in Fig.~\ref{fig:adi2}.}
\end{figure}

To investigate the effect of the configuration change on quadrupole moment, the diagonal components $Q_{mm}$ (m can be hole or particle state) of the quadrupole operator are shown in Fig.~\ref{fig:adi3} for the six specific orbits (four for neutron and two 
for proton) responsible for level crossings at the gaps A, B and C. 
A difference of the quadrupole moment between the occupied and unoccupied orbits at the gap A turned out to be $Q_{(++)_a}-Q_{(--)_a}$$\sim$18.16 ${\textrm {fm}^2}$, which is understood from Fig.~\ref{fig:adi3}(a).
Including the spin degeneracy, the second contribution is estimated as $\Delta q^{(2)}_n$$\sim$ 36.32 ${\textrm {fm}}^2$, which corresponds to 79.96\% of the neutron effect $\Delta q_n$ at the gap A. 
On the other hand, $\Delta q^{(2)}_p$=0.0  because there is no configuration change in the proton system.

The remaining contribution, i.e.,  $\Delta q_n^{(3)}$$\equiv$$\Delta q_n - \Delta q_n^{(1)} - \Delta q_n^{(2)}$ and $\Delta q_p^{(3)}$$\equiv$$\Delta q_p - \Delta q_p^{(1)} - \Delta q_p^{(2)}$, is regarded as an effect coming from the rearrangement of the mean-field, which should adjust by itself (self-consistently) in accordance with a change in the configuration.
The third contribution turned out to be $\Delta q_p^{(3)}$$\sim$17.82 ${\textrm{fm}}^2$ (88.75\% of $\Delta
q_p$) and $\Delta q_n^{(3)}$$\sim$6.23 ${\textrm{fm}}^2$ (13.71\% of $\Delta q_n$) for proton and neutron systems, respectively.
Here, it should be mentioned that the large rearrangement effect $\Delta q_p^{(3)}$ of the proton-field is induced by the neutron level crossing. 

Similar analysis are carried out for the gaps B and C. For the gap B where the crossing between neutron occupied orbit $(-,-)_b$ and unoccupied orbit $(+,+)_b$ takes place, the contributions from $\Delta q_p^{(1)}$ and $\Delta q_p^{(3)}$ to $\Delta q_p$ are 10.30 ${\textrm{fm}}^2$ (37.00\%) and 17.54 ${\textrm{fm}}^2$ (63.00\%), respectively. 
The contributions to neutron field $\Delta q_n$ from $\Delta q_n^{(1)}$, $\Delta q_n^{(2)}$ and $\Delta q_n^{(3)}$  are 13.1 ${\textrm{fm}}^2$ (17.01\%), 60.58  ${\textrm{fm}}^2$ (78.59\%) and 3.4 ${\textrm{fm}}^2$ (4.40\%), respectively.
For gap C, the contributions to $\Delta q_p$ are 1.1 ${\textrm{fm}}^2$ (2.50\%), 28.1 ${\textrm{fm}}^2$ (62.23\%) and 15.9 ${\textrm{fm}}^2$ (35.27\%),
and the contribution to neutron field $\Delta q_n$ from $\Delta q_n^{(1)}$ and $\Delta q_n^{(3)}$ are 1.44 ${\textrm{fm}}^2$ (11.46\%) and 11.12 ${\textrm{fm}}^2$ (88.54\%). From the above numerical analysis, it is clear that a large amount of shape change $\Delta q^{(2)}+\Delta q^{(3)}$ is induced by the configuration change.
Namely, the shape of the mean-field in a finite system like the nucleus is strongly affected by a few orbits involved in the configuration change, and the self-consistency between the single particle wave functions and the nuclear mean-field should be taken into account appropriately.
When one intends to discuss a relation between the diabatic and adiabatic orbits, one should take account of the above discussed large gap effects.
The large gap appears not only in the nucleus $^{70}{\textrm {Ge}}$, but generally occurs when there happens a configuration change due to the single-particle level crossing.

\subsection{\label{level32} Configuration dictated constrained HF}

\begin{figure}
\epsfxsize=8.0cm
\centerline{\epsffile{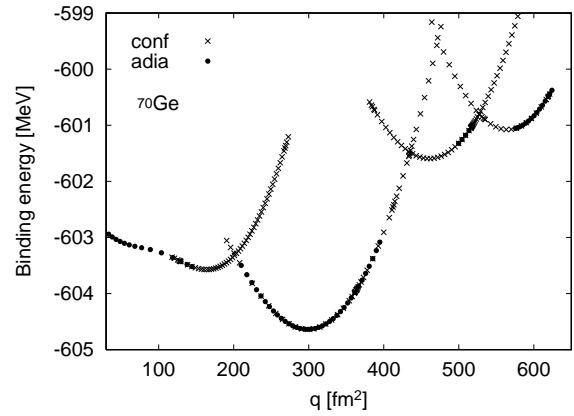}}
\caption{\label{fig:dia1} Results of the CHF potential energy curves for $^{70}{\textrm {Ge}}$. Filled circles are used for adiabatic configuration and crosses for configuration-dependent curves.}
\end{figure}

In the previous subsection, it turns out that the adiabatic potential energy curve for $^{70}{\textrm {Ge}}$ consists of four different configurations, but has only one local minimum. 
By applying the configuration dictated method, we separate the adiabatic potential energy surface (PES) into several configuration-dependent curves with different low-lying minima.
Fig.~{\ref{fig:dia1}} presents our results of the CHF potential energy curves for $^{70}{\textrm {Ge}}$. 
Here filled circles are used for the adiabatic curves which just coincides with that in Fig.~{\ref{fig:adi1}}(b), while crosses are for the configuration dictated curves. It is clearly shown that some excited minima are obtained by the configuration dictated approach, whereas only one ground state is in adiabatic PES.

The procedure to obtain the configuration dictated curves in Fig.~{\ref{fig:dia1}} are summarised as follows. 
First, the ground state of a given nucleus is obtained by optimising the three deformation parameters. 
Second, one constructs an appropriate np-nh state composed of some occupied and unoccupied orbits. 
Taking the np-nh state as an initial trial wave function for the iterative process, one may obtain an excited HF state whose configuration is dictated by RS.
Third, starting from the excited HF state, the CHF equation is solved under the maximum overlap criterion by keeping the identity of CHF states. The continuously-connected potential energy curve is obtained for a given configuration. 

The ground state of $^{70}{\textrm {Ge}}$ has the configuration [12 0, 16 2], where the last two neutrons occupy one of $g_{9/2}$ orbits. 
After obtaining the ground state, a neutron 2p-2h state composed of an excitation from the most high-lying occupied orbit $(+,+)_a$ to the second low-lying unoccupied orbit $(-,-)_a$ is constructed. 
Taking the neutron 2p-2h state as an initial trial wave function and applying the RS method, one may get an excited HF state with
excitation energy $\sim$1.1 MeV and the configuration [12 0, 18 0].
Starting from the excited HF state, the configuration-dependent curve is obtained satisfying the condition (\ref{cdm9}), a part of which just corresponds to the adiabatic solutions after the gap A shown in Fig.~\ref{fig:adi1}(b). 

Applying the same procedure, we obtain another excited HF state with excitation energy $\sim$ 2.7 MeV and the configuration [12 0, 14 4], where two neutrons are excited from the second high-lying occupied orbit $(-,-)_b$ to the most low-lying unoccupied orbits $(+,+)_b$.
The excited PES corresponds to the adiabatic solutions between gap B and C. 
The excited HF state after the adiabatic gap C, where two protons are excited from the most high-lying occupied orbit $(-,-)_c$ to the third low-lying unoccupied orbit $(+,+)_c$ and two neutrons are from $(-,-)_b$ to $(+,+)_b$, is obtained with excitation energy $\sim$ 3.6 MeV and the configuration [10 2, 14 4].

\begin{figure}
\epsfxsize=8.0cm
\centerline{\epsffile{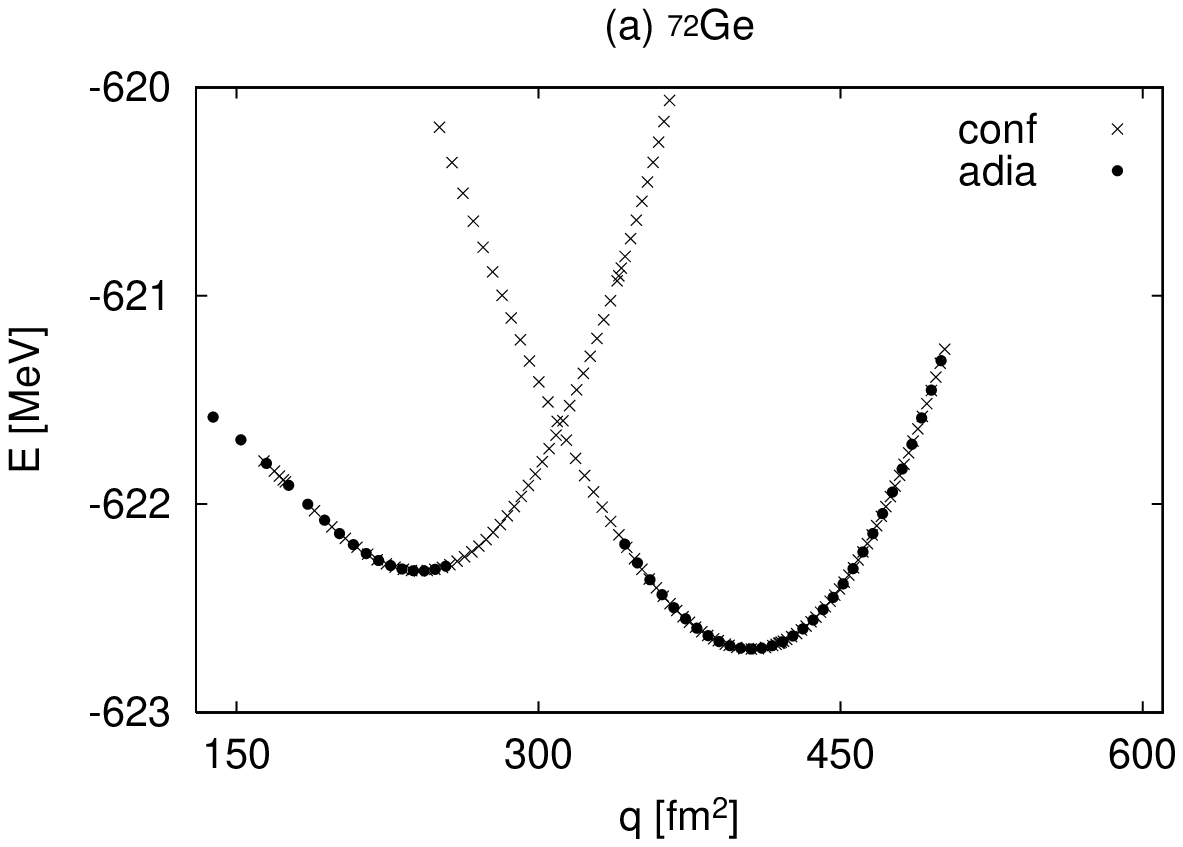}}
\centerline{\epsffile{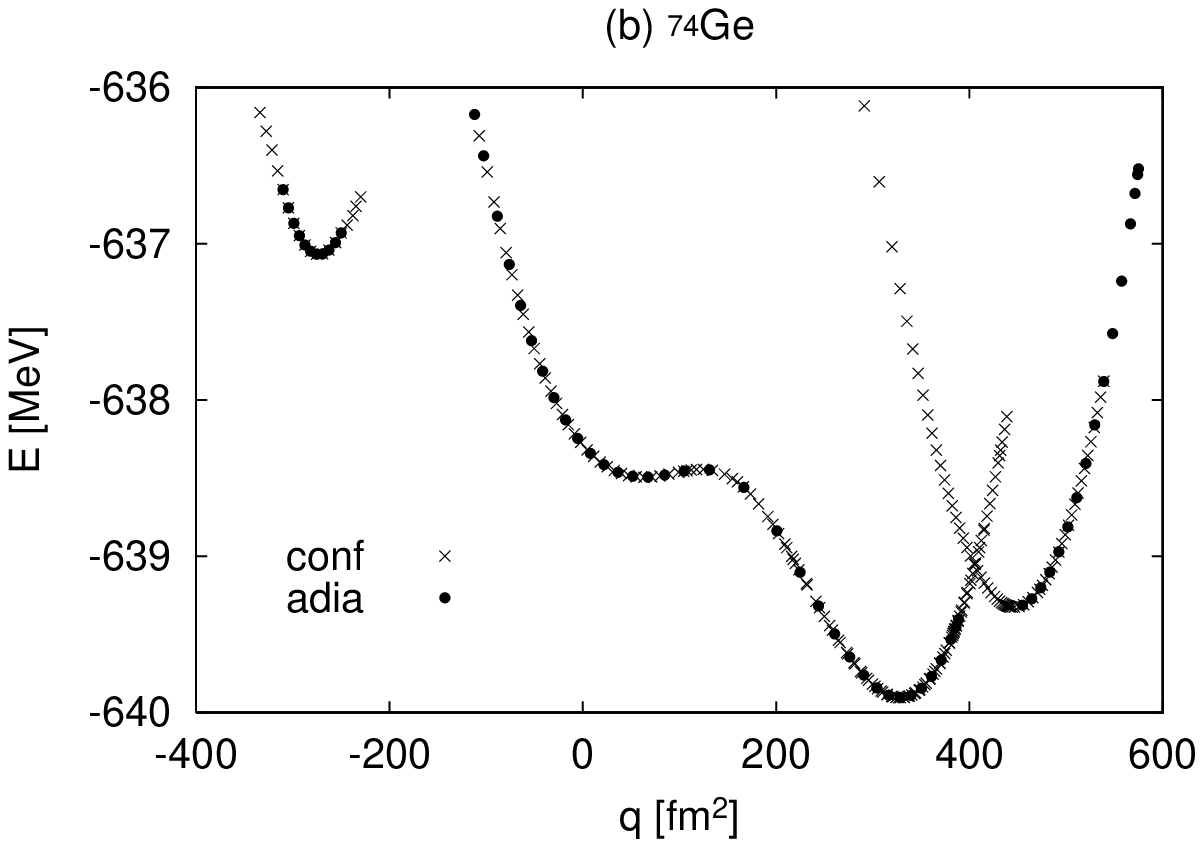}}
\centerline{\epsffile{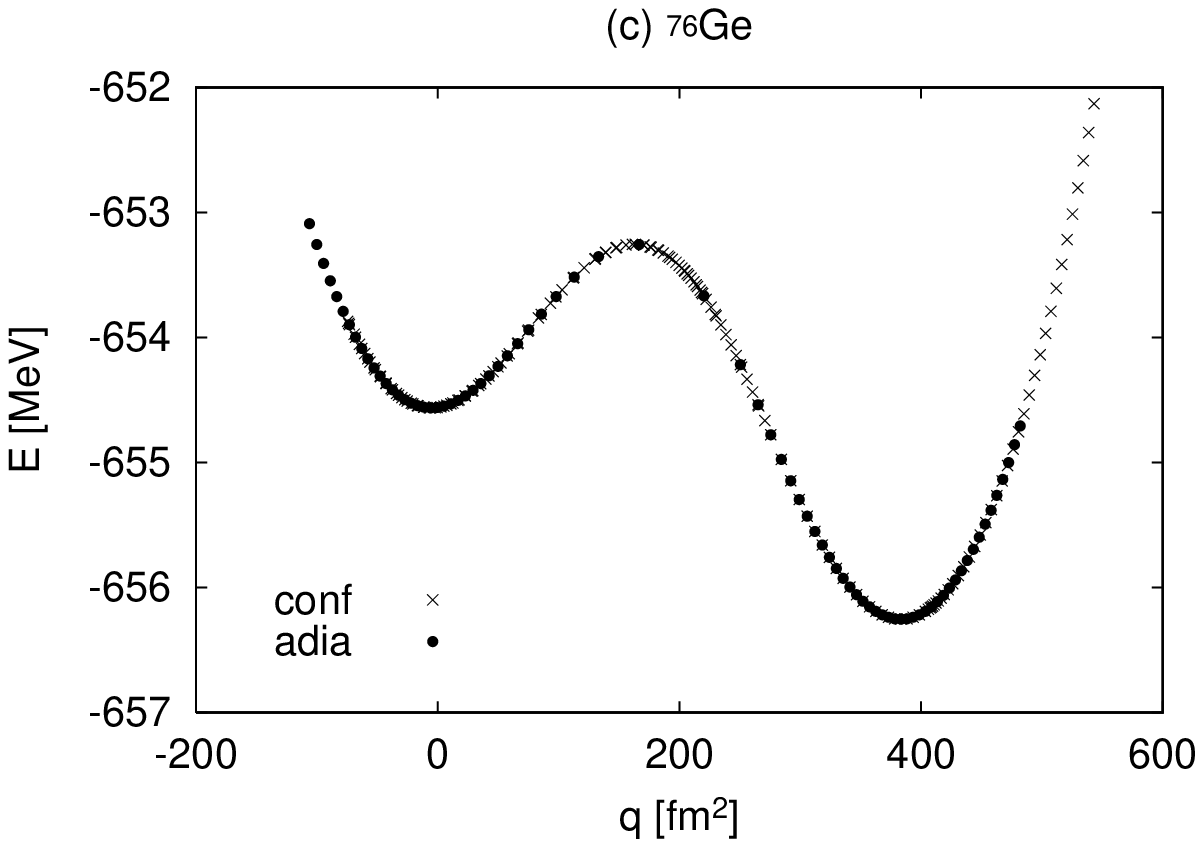}}
\caption{\label{fig:dia2} Results of the CHF potential energy curves for (a) $^{72}{\textrm {Ge}}$, (b) $^{74}{\textrm {Ge}}$ and (c) $^{76}{\textrm {Ge}}$.  Filled circles are used for adiabatic configuration and crosses for configuration-dependent curves with independent configurations.}
\end{figure}

To demonstrate systematically the discontinuity property of the adiabatic PES and the continuous property of the configuration-dependent mean-fields as a function of deformation, similar calculations are performed for $^{72}{\textrm {Ge}}$, $^{74}{\textrm {Ge}}$ and $^{76}{\textrm {Ge}}$ as shown in Fig.~\ref{fig:dia2}. 
For nucleus $^{72}{\textrm {Ge}}$, the gap in the adiabatic PES is formed by the crossing between one of the occupied $g_{9/2}$ orbits and the unoccupied $p_{1/2}$ orbit. The ground HF state is specified by configuration [12 0, 16 4] and excited HF state by [12 0, 18 2], respectively. It is observed that there are two gaps and one saddle point in the adiabatic PES of $^{74}{\textrm {Ge}}$. The gaps are also due to the crossing between the intruder orbit $g_{9/2}$ and orbit in $N_0=3$ oscillator shell, the most active orbits being $1f$ and $2p$. The saddle point is simply owing to the repulsive interaction between orbits with the same symmetries, e.g., both belonging to $g_{9/2}$. Similar saddle point is observed for $^{76}{\textrm {Ge}}$, and no gap appears in the  adiabatic PES.

For Ge isotopes within excitation energy $\sim$4 MeV, the low-lying spectra obtained by the RS method with $N_0=8$ and 12 are listed in Table~\ref{table1}, together with the available experimental data \cite{Audi,Ge1,Ge2,Ge3}. In Table~\ref{table1}, the energy differences between the excited states and ground state given by $\Delta E=[B.E.(\mbox{ground})-B.E.(\mbox{excited})]$ are listed. 

The configurations specifying the excited HF states are also listed in Tab.~\ref{table1}. 
Here, it should be mentioned that the configurations $[p_1p_2,n_1n_2]$ specifying the HF states are the same for $N_0=12$ and $N_0=8$.
As noted in Ref.~\cite{Patra}, a triaxial calculation is necessary to obtain the low-lying minima in the mass region A=60$\sim$80, since some of them are $\gamma$-soft nuclei. From our numerical calculation and experimental data, the ground and excited states of Ge isotopes are considered to have a triaxial deformation. 
Table~\ref{table1} shows that the binding energy for $N_0=12$ is about 2 MeV lower than those in $N_0=8$, while the excitation energy and deformations are quite similar for $N_0=8$ and 12. 
From the above discussion, the size of configuration space is rather sensitive for the total binding energy, but not for the excitation energy, quadrupole nor triaxial deformations. 

In this mass region, the spherical or normal deformation are characterised by a configuration where the $N_0=3$ oscillator shell is occupied, whereas the larger deformation by a configuration with the occupied $N_0=4$ ($g_{9/2}$ orbits) oscillator shell \cite{JL92,WN85,Patra}. For example, the first excited state in $^{70}{\textrm {Ge}}$ is characterised by a neutron $2p$-$2h$ state where two neutrons in the deformation-driving orbit $g_{9/2}$ are excited to the orbit in $N_0=3$ oscillator shell. Consequently, the first excited state has smaller quadrupole deformation $\beta_2$ (0.17) than the ground state (0.24). The second excited state is two neutrons
excitation from $N_0=3$ shell to the intruder orbit $g_{9/2}$, resulting in the relatively larger $\beta_2$ deformation (0.35). 
Since the third (fourth) excited state has similar configuration with the ground (second excited) state, the third (fourth) excited state has the similar $\beta_2$ deformation with the ground (second excited) state. 
Upon the second excited state, another two protons are excited to the intruder orbit $g_{9/2}$, forming the fifth excited state with superdeformed character ($\beta_2=0.42$).
 Deformation character of low-lying minima for $^{72}{\textrm {Ge}}$, $^{74}{\textrm {Ge}}$ and $^{76}{\textrm {Ge}}$ shows the similar situation with those in $^{70}{\textrm {Ge}}$ (see Table~\ref{table1}).
It is clear from Tab.~\ref{table1} that our calculation indicates some low-lying excited states in this mass region. 
An experimental study on these shape coexisting phenomena would be of special interest.

\subsection{\label{level33}  Effect of pairing correlation }
\begin{figure}
\epsfxsize=8.0cm
\centerline{\epsffile{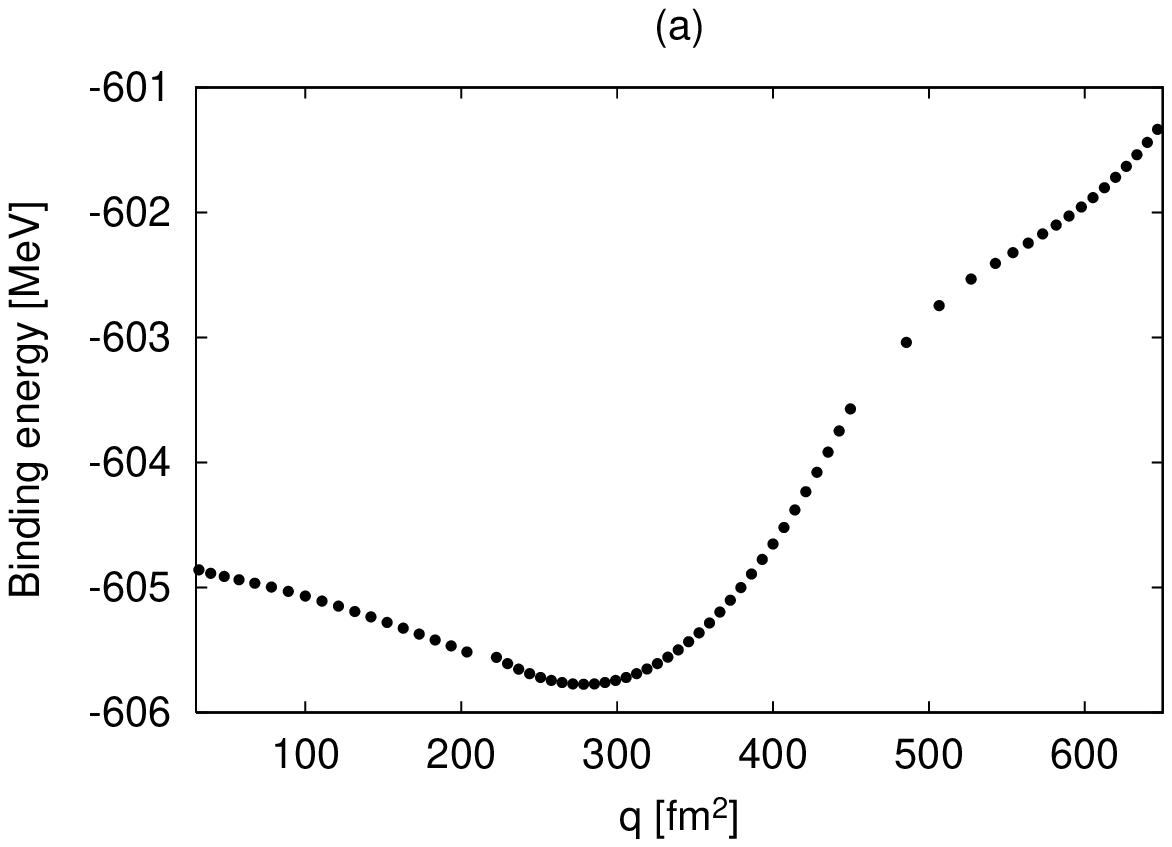}}
\centerline{\epsffile{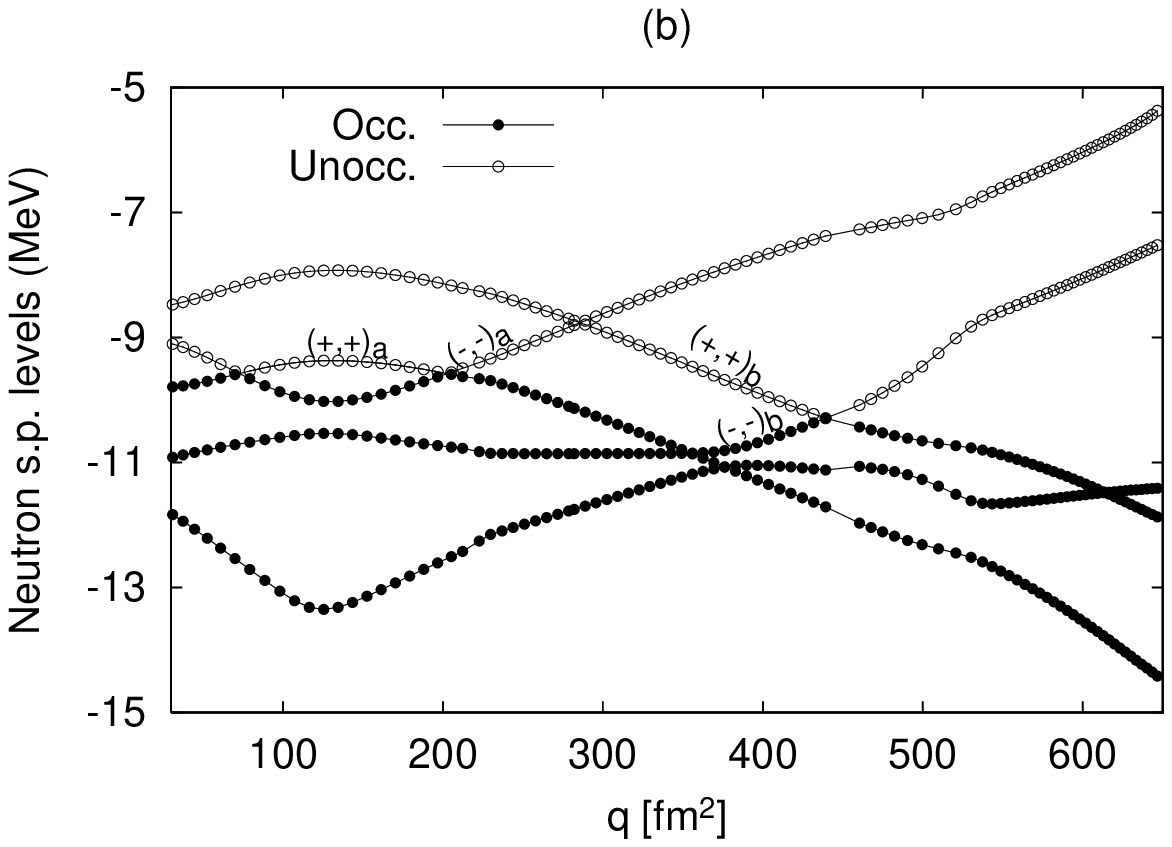}}
\caption{\label{fig:hfb} Adiabatic CHFB for
$^{70}{\textrm {Ge}}$: (a) binding energy as a function of the quadrupole moment; 
(b) neutron single-particle orbits in canonical basis. Occ. and Unocc. stand for occupied and unoccupied orbits, respectively. The corresponding orbits are denoted by the same symbols as CHF calculation.}
\end{figure}

We have also performed the adiabatic CHFB calculation with Gogny force to include the pairing correlation into the mean-field self-consistently. Figure~\ref{fig:hfb}(a) shows the lowest HFB potential energy surface (PES) for $^{70}{\textrm {Ge}}$, and the properties of the HFB ground state for Ge isotopes are listed in Tab.~\ref{table1}. 
From this table, one may observe the binding energy increases 0.59 to 1.34 MeV after having included the pairing correlation into the mean-field.
In so far as the $\beta_2$ and $\gamma$ deformation are concerned, however, our numerical calculation does not show any substantial differences between the HF and HFB states, which is recognised from Tab.~\ref{table1}.
Since our triaxial deformed results both in the HF and HFB calculations well reproduce the available experimental data from nucleus $^{70}{\textrm {Ge}}$ to $^{76}{\textrm {Ge}}$, the pairing correlation is not decisive for the ground state properties except for the binding energy.

To study the effects of the pairing on the quadrupole deformation more deeply, in Tab.~\ref{table2}, numerical values of $\langle{\hat Q}_{20}\rangle$ and $\langle{\hat Q}_{22}\rangle$ for Ge isotopes are listed for both the HF and HFB calculations.
From Tabs.~\ref{table1} and ~\ref{table2}, one may state that the value $\langle{\hat Q}_{20}\rangle$ changes strongly depending on the neutron number, whereas that of $\beta_2$ does not. 
In this sense, $\beta_2$ and $\gamma$ may not be regarded to be good quantities to specify the shape of the nucleus in this mass region.
A comparison between the HF and HFB calculations in Tab.~\ref{table2} tells that the values of $\langle{\hat Q}_{20}\rangle$ and $\langle{\hat Q}_{22}\rangle$ generally become small (favor the spherical shape) when one includes the pairing correlation into the mean-field, except for a case in the value $\langle{\hat Q}_{20}\rangle$ of $^{74}{\textrm {Ge}}$ where the microscopic shell structure plays a role. 

In order to study what happens in the level crossing region after having included the pairing correlation, the neutron single-particle energies of the canonical basis for $^{70}{\textrm {Ge}}$ in the adiabatic CHFB calculation are shown in Fig.~\ref{fig:hfb}(b).
From this figure, it is observed that the configuration change induced by the crossing between two specific orbits still occurs in the CHFB calculation, even though the gaps, i.e., the missing regions in the adiabatic PES are not so distinct in comparison with the case of CHF in Fig.~\ref{fig:adi1}.
A smallness of the missing region may be understood by the following reason: two configurations having different shape are mixed up by the pairing correlation, and the $uv$-factor introduced by the BCS theory makes a concept of the configuration obscure. 
However, it turned out to be very difficult to get the excited HFB states when one applies the configuration dictated CHFB method, unlike the CHF case shown in Fig.~\ref{fig:dia1}. (This is a reason why only the ground state properties in HFB calculation are presented in Table~\ref{table1}).
In the case of CHF, the gap appeared in Fig.~\ref{fig:adi1} does not indicate any difficulty of the mean-field theory, but is caused by the artificial adiabatic assumption put by hand.
Applying the configuration dictated CHF method, one usually finds a continuation of the PES in the gap region, and obtains many CHF states with different configurations as discussed in the previous subsection.
On the other hand, in the case of CHFB, the missing region remains no matter how much efforts we made in applying the configuration dictated method.
From our numerical calculations, we may deduce the following conclusions: (1) the CHFB state can not be well characterised by the configuration defined in the canonical basis, and (2) the lowest PES in the CHFB is more fragile than that of the CHF in the level crossing region.

Here, it should be mentioned that the configuration dictated CHF calculation also meets a difficulty of poor convergence or even non-convergence, when a pair of avoided crossing orbits comes close with each other.
From our analytic and numerical studies \cite{Guo2}, it turned out that the CHF iterations diverge when the quantum fluctuations coming from the two-body residual interaction and quadrupole deformation become comparable with an energy difference between two avoided crossing orbits in the CHF one-body potential.
The above stated difficulty is related to an applicability of the mean-field theory, when some part of two-body interaction could not be approximated by one-body potential successfully, but always remains there during the iterations. 
After including the pairing correlations, this situation might become more complicated, because there are competitions not only between the ph type two-body residual interaction and the HF potential, between the pp type two-body residual interaction and the pairing potential, but also their cross effects. 
These interesting problem will be discussed elsewhere \cite{Guo2}.

\section{\label{level4} CONCLUDING REMARKS}
In the present work, the adiabatic and configuration dictated CHF(B) calculations are performed for Ge isotopes, and the following three points are discussed. 

First: by using the adiabatic CHF calculation, it is shown that
 the shape of the CHF mean-field in a finite system is strongly affected by only a few orbits involved in the configuration change.
This situation appears not for the specific nucleus, but generally occurs when there happens a configuration change in association with the single particle level crossing. 
One should thus carefully take account of the differences between the mean-fields accompanied with the diabatic orbits and with the adiabatic orbits.
 
Second:  the configuration dictated CHF method for Ge isotopes gives the low-lying spectra as well as the low-lying PESs, which nicely reproduce the triaxial deformed character of Ge isotopes. As far as the ground state properties of Ge isotopes are concerned, the pairing correlations are not decisive except for the binding energies.
Since our numerical calculation indicates some additional low-lying excited $0^+$ states, an experimental study on these shape coexisting states in this mass region would be interesting.

Last: based on our numerical calculations,  the PES in the CHFB is expected to be more fragile than that in the CHF in the level crossing region.
Further study on the effects of pairing correlation near the level crossing region is needed in terms of the competitions between the two-body residual interaction and the one-body HFB potential, and between the average quantities like the $uv$-factors and their quantum fluctuations.

\begin{acknowledgments}
One of us (L. Guo) thanks Dr. Y. Hashimoto, Dr. T. Tanaka, Dr. K. Iwasawa, Dr. J. Meng for their valuable discussions and useful comments. This work was supported in part by the Japan Society for the Promotion of Science (JSPS) and the China National Natural Science Foundation (CNSF) as the bilateral program between Japan and China. Another author (E. G. Zhao) acknowledges the support by Natural Science
Foundation of China under Grant No. 10375001,  the China  Major Basic
Research Development Program under Grant No. G2000-0774-07, the Knowledge
Innovation Project of the Chinese Academy of Sciences under Grant No.
KJCX2-SW-N02.
\end{acknowledgments}

\bibliography{NPA1}

\newpage

\begingroup
\squeezetable
\begin{table} 
\caption{\label{table1} The triaxial deformed low-lying spectra by RS method for the binding energy (BE), quadrupole deformation parameter ($\beta_2$),
triaxial deformation ($\gamma$), and the difference between the ground state and excited state energies ($\Delta E$). Results with $N_0=8$ and 12 are given for comparison. The configuration $[p_1p_2,n_1n_2]$ for $N_0=8$ is same to the case in $N_0=12$ and hence are not listed here.
The ground state properties are presented in HFB calculation.
The experimental data for binding energy \cite{Audi}, quadrupole and triaxial deformation parameters \cite{Ge1,Ge2,Ge3} 
are also listed. The energy is in MeV and triaxial deformation is in degree.
}
\begin{ruledtabular}
\begin{tabular}{cccccccccccccccc}
Nucleus & \multicolumn{4}{c}{$N_0=8$ (HF)} & \multicolumn{5}{c}{$N_0=12$ (HF)}  & \multicolumn{3}{c}{$N_0=12$ (HFB)}  & \multicolumn{3}{c}{Expt.} \\
        & BE & $\beta_2$ & $\gamma$ & $\Delta E$ & BE & $\beta_2$ & $\gamma$ & $\Delta E$ & $[p_1p_2,n_1n_2]$ & BE & $\beta_2$ & $\gamma$ & BE & $\beta_2$ & $\gamma$  \\
\hline
$^{70}{\textrm {Ge}}$  &  604.64 & 0.25 & 35.2  &      &  606.59 & 0.24 & 33.9 &          & [12 0, 16 2]  & 607.66 & 0.24 & 33.4 & 610.51 & 0.23 & 30.8 \\
                       &  603.57 & 0.16 & 42.1  & 1.07 &  605.44 & 0.17 & 40.6 & 1.15     & [12 0, 18 0]  &        &      &      &        &      &      \\
                       &  601.97 & 0.32 & 33.5  & 2.67 &  603.90 & 0.35 & 30.9 & 2.69     & [12 0, 14 4]  &        &      &      &        &      &      \\
                       &  601.67 & 0.24 & 25.2  & 2.97 &  603.50 & 0.25 & 23.6 & 3.09     & [12 0, 16 2]  &        &      &      &        &      &      \\
                       &  601.18 & 0.36 & 8.01  & 3.46 &  603.00 & 0.35 & 6.70 & 3.59     & [12 0, 14 4]  &        &      &      &        &      &      \\
                       &  601.01 & 0.40 & 0.33  & 3.63 &  602.87 & 0.42 & 0.18 & 3.72     & [10 2, 14 4]  &        &      &      &        &      &      \\
$^{72}{\textrm {Ge}}$  &  622.69 & 0.29 & 22.6  &      &  624.75 & 0.28 & 23.8 &          & [12 0, 16 4]  & 626.09 & 0.25 & 28.5 & 628.68 & 0.25 & 33.6 \\
                       &  622.32 & 0.21 & 38.2  & 0.37 &  624.32 & 0.23 & 37.6 & 0.43     & [12 0, 18 2]  &        &      &      &        &      &      \\
                       &  621.88 & 0.21 & 40.7  & 0.81 &  623.92 & 0.21 & 39.1 & 0.83     & [12 0, 18 2]  &        &      &      &        &      &      \\
                       &  620.53 & 0.28 & 28.7  & 2.16 &  622.47 & 0.27 & 27.4 & 2.28     & [12 0, 18 2]  &        &      &      &        &      &      \\
                       &  619.98 & 0.36 & 15.6  & 2.71 &  621.97 & 0.38 & 13.6 & 2.78     & [10 2, 16 4]  &        &      &      &        &      &      \\
                       &  619.61 & 0.18 & 54.2  & 3.08 &  621.62 & 0.17 & 56.2 & 3.13     & [12 0, 18 2]  &        &      &      &        &      &      \\
$^{74}{\textrm {Ge}}$  &  639.91 & 0.24 & 26.5  &      &  642.06 & 0.24 & 27.6 &          & [12 0, 18 4]  & 643.31 & 0.25 & 26.9 & 645.66 & 0.28 & 25.8 \\
                       &  639.33 & 0.31 & 23.5  & 0.58 &  641.46 & 0.31 & 22.6 & 0.60     & [12 0, 16 6]  &        &      &      &        &      &      \\
                       &  638.49 & 0.21 & 18.8  & 1.42 &  640.59 & 0.23 & 17.8 & 1.47     & [12 0, 18 4]  &        &      &      &        &      &       \\
                       &  638.08 & 0.22 & 31.8  & 1.83 &  640.18 & 0.22 & 30.6 & 1.88     & [12 0, 18 4]  &        &      &      &        &      &       \\
                       &  637.38 & 0.14 & 24.5  & 2.53 &  639.40 & 0.15 & 24.0 & 2.66     & [12 0, 20 2]  &        &      &      &        &      &       \\
                       &  636.23 & 0.21 & 46.5  & 3.68 &  638.31 & 0.23 & 45.8 & 3.75     & [12 0, 18 4]  &        &      &      &        &      &       \\
$^{76}{\textrm {Ge}}$  &  656.25 & 0.26 & 25.7  &      &  658.55 & 0.25 & 25.0 &          & [12 0, 18 6]  & 659.14 & 0.24 & 24.5 & 661.59 & 0.26 & 28.9   \\
                       &  654.56 & 0.24 & 30.8  & 1.69 &  656.83 & 0.24 & 31.6 &1.72      & [12 0, 18 6]  &        &      &      &        &      &        \\
                       &  654.38 & 0.18 & 12.7  & 1.87 &  656.59 & 0.17 & 14.2 &1.96      & [12 0, 20 4]  &        &      &      &        &      &        \\
                       &  653.32 & 0.23 & 37.6  & 2.93 &  655.57 & 0.22 & 38.9 & 2.98     & [12 0, 18 6]  &        &      &      &        &      &        \\
                       &  652.85 & 0.31 & 29.5  & 3.40 &  655.07 & 0.33 & 28.6 & 3.48     & [12 0, 16 8]  &        &      &      &        &      &         \\
                       &  651.76 & 0.23 & 35.1  & 4.49 &  653.93 & 0.24 & 36.8 & 4.62     & [12 0, 18 6]  &        &      &      &        &      &         \\
\end{tabular}
\end{ruledtabular}
\end{table}
\endgroup

\begin{table}
\caption{\label{table2} Expectation values of quadrupole operators $\hat Q_{20}$ and $\hat Q_{22}$ for the ground state of Ge isotopes
in HF and HFB calculations. $\langle\hat Q_{20}\rangle$ and $\langle\hat Q_{22}\rangle$ are in ${\textrm {fm}^2}$.}
\begin{ruledtabular}
\begin{tabular}{ccccc}
Nucleus & HF &  & HFB &  \\
        &  $\langle\hat Q_{20}\rangle$  & $\langle\hat Q_{22}\rangle$ & $\langle\hat Q_{20}\rangle$ & $\langle\hat Q_{22}\rangle$ \\
\hline
$^{70}{\textrm {Ge}}$  &  298.636  & 116.276  &  278.566 &  105.906  \\
$^{72}{\textrm {Ge}}$  &  405.678  & 103.241  &  324.727 &  101.721   \\
$^{74}{\textrm {Ge}}$  &  328.358  &  99.223  &  336.873 &   98.719   \\
$^{76}{\textrm {Ge}}$  &  383.277  & 103.348  &  354.340 &   93.343   
\end{tabular} 
\end{ruledtabular}
\end{table}

\end{document}